\documentclass[reprint,showpacs,preprintnumbers,amsmath,amssymb,prc,floatfix]{revtex4-1}
\usepackage{color}
\usepackage{graphicx}
\usepackage{dcolumn}
\usepackage{bm}
\usepackage{amstext}
\usepackage{verbatim}
\usepackage[colorlinks,citecolor=blue,linkcolor=red,anchorcolor=blue,filecolor=blue,urlcolor=blue]{
hyperref}

\begin{document}

\title{Consistent Skyrme parametrizations constrained by GW170817}

\author{O. Louren\c{c}o$^1$, M. Dutra$^1$, C. H. Lenzi$^1$, S. K. Biswal$^2$, M. 
Bhuyan$^3$, and D. P. Menezes$^4$}  

\affiliation{
\mbox{$^1$Departamento de F\'isica, Instituto Tecnol\'ogico da Aeron\'autica, DCTA, 12228-900, 
S\~ao Jos\'e dos Campos, SP, Brazil} \\
\mbox{$^2$Key Laboratory of Theoretical Physics, Institute of Theoretical Physics, Chinese Academy 
of Sciences, Beijing 100190, China} \\
\mbox{$^3$ Department of Physics, Faculty of Science, University of Malaya, Kuala Lumpur 50603, 
Malaysia}\\
\mbox{$^4$Depto de F\'{\i}sica - CFM - Universidade Federal de Santa Catarina, Florian\'opolis - SC 
- CP. 476 - CEP 88.040 - 900 - Brazil}}

\begin{abstract}
The high-density behavior of the stellar matter composed of nucleons and leptons 
under 
$\beta$~equilibrium and charge neutrality conditions is studied with the Skyrme parametrizations 
shown to be consistent (CSkP) with the nuclear matter, pure neutron matter, symmetry energy and its 
derivatives in a set of $11$ constraints [Dutra {\it et al.}, Phys. Rev. C 85, 035201 (2012)]. The 
predictions of these parametrizations on the tidal deformabilities related to the GW170817 event are 
also examined. The CSkP that produce massive neutron stars give a range of 
$11.86~\mbox{km} \leqslant R_{1.4} \leqslant 12.55~\mbox{km}$ for the canonical 
star radius, in agreement with other theoretical predictions. It is shown that the CSkP are 
compatible with the region of masses and radii obtained from the analysis of recent data from LIGO 
and Virgo Collaboration (LVC). A correlation between dimensionless tidal deformability and radius of 
the canonical star is found, namely, $\Lambda_{1.4} \approx 
3.16\times10^{-6}R_{1.4}^{7.35}$, with results for the CSkP compatible with the recent range of 
$\Lambda_{1.4}=190_{-120}^{+390}$ from LVC. An analysis of the $\Lambda_1\times\Lambda_2$ graph 
shows that all the CSkP are compatible with the recent bounds obtained by LVC. Finally, the 
universal correlation between the moment of inertia and the deformability of a neutron star, named 
as the \mbox{$I$-Love} relation, is verified for the CSkP, that are also shown to be consistent 
with the prediction for the moment of inertia of the \mbox{PSR J0737-3039} primary component pulsar.
\end{abstract}
\maketitle

\section{Introduction}
\label{intro}

Neutron stars are incredible natural laboratories for the study of nuclear matter at extreme 
conditions of isospin asymmetry and density ($\rho$)~\cite{latt04,ozel06}. The properties of 
nuclear matter at such high densities are mostly governed by the equation(s) of state (EOS), which 
correlates pressure ($p$), energy density ($\epsilon$) and other thermodynamical quantities. From 
the terrestrial experiments, nuclear matter properties are mostly constrained up to saturation 
density, $\rho_0\approx0.15$~fm$^{-3}\approx2.8\times10^{14}$~g/cm$^3$~\cite{tsan12,bald16,latt16,
 oert17}. The EOS correlating $p$, $\epsilon$ and $\rho$ is the sole ingredient to determine the 
relationship between the mass and radius of a neutron star by using the Tolman-Oppenheimer-Volkoff 
equations~\cite{tov39,tov39a}. It also plays a vital role in determining other star properties such 
as the moment of inertia and tidal deformability~\cite{tanj10,phil18}. These last two quantities 
are shown to be correlated according to the so called $I$-Love relation~\cite{phil18,science}. 
The measurements of the neutron star spin, radius, and gravitational redshift provide weak 
constraints on the EOS as these measurements depend on the detailed modeling of the radiation 
mechanism and are subjected to a lot of systematic errors~\cite{latt07,joce09}. However,the recent 
observation of the gravitational waves (GW) emission from the first binary neutron stars merger 
event, GW170817, provided new expectations to constrain the EOS in more efficient 
ways~\cite{ligo17,ligo18}.

Since 2015, the observation of the GW emission from the binary compact objects, by 
LIGO~\cite{aasi15} and Virgo~\cite{acer15} collaborations, opened a platform to study the GW and 
related physics in more adequate ways. The GW170817 event, observed on 17 August 2017, has a special 
importance in nuclear physics since it consists of the emergence of GW from a binary system of neutron 
stars. It coincides with the detection of the $\gamma$-ray burst GRB170817~\cite{abbo17,gold17} and 
the components were verified as neutron stars by various electromagnetic spectrum 
observations~\cite{abbo17a,coul17,troj17,hagg17,hall17}. Hence, the GW170817 offers an opportunity to constrain the EOS from the tidal deformability data~\cite{bhar17,tanj10,hind08,thib09,tayl09}, which establishes a relation 
between the internal structure of the neutron star and the emitted GW.

In the present context, we use the Skyrme model~\cite{skyr61,vautherin,bend03,ston07} in order to 
explore the possible constraints on the EOS by the observation of the GW170817 event. In the work 
of Ref.~\cite{dutra12}, the authors have studied the nuclear matter characteristics of symmetric 
and asymmetric matter at saturation as well as at high densities by using $240$ parametrizations of 
the Skyrme energy density functional. Following this work, it was observed that $16$ 
parametrizations, namely, GSkI \cite{agrawal2006}, GSkII~\cite{agrawal2006}, KDE0v1 
\cite{agrawal2005}, LNS \cite{cao2006}, MSL0~\cite{chen2010}, NRAPR~\cite{steiner2005}, Ska25s20 
\cite{private2}, Ska35s20~\cite{private2}, SKRA~\cite{rashdan2000}, Skxs20 \cite{brown2007}, 
SQMC650~\cite{guichon2006}, SQMC700 \cite{guichon2006}, SkT1~\cite{tondeur1984,stone2003}, 
SkT2~\cite{tondeur1984,stone2003}, SkT3~\cite{tondeur1984,stone2003} and 
\mbox{SV-sym32}~\cite{klupfel2009}, satisfy all the chosen $11$ constraints analyzed in 
Ref.~\cite{dutra12} from symmetric nuclear matter, pure neutron matter, and a mixture of both 
related with the symmetry energy and its derivatives~\cite{bianca}. This set was named as Consistent 
Skyrme Parametrizations (CSkP), which is used in the present manuscript. These parametrizations 
offer a predictive power starting from sub-saturation density to very high density at very high 
isospin asymmetry, what has motivated us to analyze the stellar matter behavior for the CSkP, in 
particular, the tidal deformability related to the GW170817 event. 

We point out that the constraints analyzed in Ref.~\cite{dutra12} were chosen by 
collecting previous existing constraints in the literature at the time. In that sense, they are not 
unique and can be subject to improvements and/or updates. Actually,
in subsequent works, relativistc mean-field models were  analyzed by a set of updated 
constraints ~\cite{rmf,rmfdef} in comparison with those used in Ref.~\cite{dutra12}. However, 
we also remark here that the main purpose of the paper in Ref.~\cite{dutra12} was to establish some 
general criteria and analyse the models according to them. The purpose of the present paper is not 
to justify the criteria chosen in Ref.~\cite{dutra12}, but to use the models consistent with all 
those chosen constraints to verify whether (or not) they could also satisfy the new 
constraints imposed by the GW170817 event.

We try to correlate the tidal deformability of the canonical neutron star 
($\Lambda_{1.4}$) and the corresponding radius ($R_{1.4}$) for the CSkP by addressing a transparent 
relation between $\Lambda_{1.4}$ and $R_{1.4}$ as a power law. Usually, the proportionality 
relation $\Lambda \propto R^5$, which is based on the definition $\Lambda = (2/3)k_2 (R/M)^5$, with 
$M$ being the neutron star mass, is cited in the literature. It is worth noticing that this 
proportionality is not exact since the Love number $k_2$ depends on the radius $R$ through a 
complicated second order differential equation. In recent studies, various relations between the 
$\Lambda_{1.4}$ and $R_{1.4}$ are obtained with different models, like the Skyrme~\cite{malik18} and 
relativistic mean-field~\cite{fatt18} ones. Here, we study this correlation with CSkP. The 
predictions of the CSkP regarding the values for $\Lambda_{1.4}$ and the tidal deformabilities of 
the binary neutron star system, namely, $\Lambda_1$ and $\Lambda_2$ are also presented. The 
verification of the $I$-Love relation, and the predictions for the dimensionless moment of inertia 
of the \mbox{PSR J0737-3039} primary component pulsar, obtained from the CSkP, are also performed.

This manuscript is organized as follows. In Sec.~\ref{theory}, we briefly outline the theoretical 
formalism of the Skyrme model in nuclear and neutron star matter. In Sec.~\ref{result}, we discuss 
the predictions of CSkP concerning the recent constraints obtained from the GW170817 event and 
verify the $I$-Love relation. Special attention is given to the tidal deformability of the neutron 
stars binary system. We conclude the manuscript with a brief summary in Sec.~\ref{summary}.

\section{Theoretical Formalism}
\label{theory}

\subsection{Infinite nuclear matter}

In the following, we mention the EOS used in this work related to the Skyrme model at zero 
temperature. The energy density of infinite nuclear matter, defined in terms of the density and proton fraction, is written as~\cite{dutra12}
\begin{align}
\epsilon(\rho,y_p) &= 
\frac{3\hbar^2}{10M_{\rm nuc}}\left(\frac{3\pi^2}{2}\right)^{2/3}\rho^{5/3}H_{5/3}(y_p)
\nonumber \\ 
&+ \frac{t_0}{8}\rho^2[2(x_0+2)-(2x_0+1)H_2(y_p)] 
\nonumber \\
&+ \frac{1}{48}\sum_{i=1}^{3}t_{3i}\rho^{\sigma_{i}+2}[2(x_{3i}+2)-(2x_{3i}+1)H_2(y_p)]
\nonumber \\
&+\frac{3}{40}\left(\frac{3\pi^2}{2}\right)^{2/3}\rho^{8/3}[aH_{5/3}(y_p) + bH_{8/3}(y_p)],
\label{edsk}
\end{align}
with
\begin{eqnarray}
a &=& t_1(x_1+2)+t_2(x_2+2),
\\
b &=& \frac{1}{2}\left[t_2(2x_2+1)-t_1(2x_1+1)\right],
\end{eqnarray}
and
\begin{eqnarray}
H_l(y_p)&=&2^{l-1}[y_p^l+(1-y_p)^l],
\end{eqnarray}
where $y_p=\rho_p/\rho$ is the proton fraction, and $M_{\rm nuc}c^2=939$~MeV is the nucleon rest 
mass, assumed equal for protons and neutrons. A particular parametrization is defined by a specific 
set of the following free parameters: $x_0$, $x_1$, $x_2$, $x_{31}$, $x_{32}$, $x_{33}$, $t_0$, 
$t_1$, $t_2$, $t_{31}$, $t_{32}$, $t_{33}$, $\sigma_1$, $\sigma_2$, and $\sigma_3$.

From Eq.~(\ref{edsk}), one can construct the pressure of the model as
\begin{align}
&p(\rho,y_p) = \rho^{2}\frac{\partial(\mathcal{E}/\rho)}{\partial\rho} =  
\frac{\hbar^2}{5M_{\rm nuc}}\left(\frac{3\pi^2}{2} \right)^{2/3}\rho^{5/3}H_{5/3}(y_p) 
\nonumber \\
&+\frac{t_0}{8}\rho^2[2(x_0+2)-(2x_0+1)H_2(y_p)]
\nonumber \\
&+\frac{1}{48}\sum_{i=1}^{3}t_{3i}(\sigma_i+1)\rho^{\sigma_i+2}[2(x_{3i}+2)-(2x_{3i}
+1)H_2(y_p)]
\nonumber \\
&+\frac{1}{8}\left(\frac{3\pi^2}{2}\right)^{2/3}\rho^{8/3}[aH_{5/3}(y_p) + bH_{8/3}(y_p)].
\label{prsk}
\end{align}
Notice that pressure and energy density are given in units of MeV/fm$^3$ and can be 
converted into units of fm$^{-4}$ by using the following conversion factor: 
$1$~fm$^{-4}=197.33$~MeV/fm$^3$. The nucleon chemical potential is given by
\begin{align}
&\mu_q(\rho,y_p) = \frac{\partial\epsilon}{\partial\rho_q} =
\frac{\hbar^2}{2M_{\rm nuc}}\left(\frac{3\pi^2}{2}\right)^{2/3}\rho^{2/3}H_{5/3}(y_p)
\nonumber\\
&+ \frac{1}{5}\left(\frac{3\pi^2}{2}\right)^{2/3}\rho^{5/3}[aH_{5/3}(y_p) + bH_{8/3}(y_p)]
\nonumber\\
&+\frac{t_0}{4}\rho[2(x_0+2)-(2x_0+1)H_2(y_p)]
\nonumber\\
&+ 
\frac{1}{48}\sum_{i=1}^{3}t_{3i}(\sigma_i+2)\rho^{\sigma_i+1}[2(x_{3i}+2)-(2x_{3i}
+1)H_2(y_p)]
\nonumber\\
&\pm \frac{1}{2}\left[1 \mp 
(2y_p-1)\right]\left\{\frac{3\hbar^2}{10M_{\rm nuc}}\left(\frac{3\pi^2}{2}\right)^{2/3} 
\rho^{2/3}H'_{5/3}(y_p) \right.
\nonumber\\
&\left.-\frac{t_0}{8}\rho(2x_0+1)H'_2(y_p) 
-\frac{1}{48}\sum_{i=1}^{3}t_{3i}\rho^{\sigma_i+1}(2x_{3i} +1)H'_2(y_p) \right.
\nonumber\\
&+\left. \frac{3}{40} 
\left(\frac{3\pi^2}{2}\right)^{2/3}\rho^{5/3}[aH'_{5/3}(y_p) + bH'_{8/3}(y_p)]\right\},
\label{muqsk}
\end{align}
where $q=p,n$ stands for protons and neutrons, respectively. Here one also has that 
$H'_l(y_p)=dH_l/dy_p$.

\subsection{Neutron star matter}

For a correct treatment of the stellar matter, one needs to implement charge neutrality and 
$\beta$-equilibrium conditions under the weak processes, $n\rightarrow p + e^- + \bar{\nu}_e$, and 
its inverse process $p+ e^-\rightarrow n + \nu_e$. For densities in which $\mu_e$ exceeds the muon 
mass, the reactions $e^-\rightarrow \mu^- + \nu_e + \bar{\nu}_\mu$, $p + \mu^-\rightarrow n + 
\nu_\mu$, and $n\rightarrow p + \mu^- + \bar{\nu}_\mu$ energetically favor the emergence of muons. 
Here, we consider that neutrinos are able to escape the star due to their extremely small 
cross-sections at zero temperature. By taking these assumptions into account, we can write the total 
energy density and pressure of the stellar system for the Skyrme model, respectively, as
\begin{align}
\mathcal{E}(\rho,\rho_e,y_p) &= \epsilon(\rho,y_p) + \frac{\mu_e^4(\rho_e)}{4\pi^2\hbar^3 c^3}
+ M_{\rm nuc}c^2\rho
\nonumber\\
&+ 
\frac{1}{\pi^2\hbar^3 c^3}\int_0^{\sqrt{\mu_\mu^2(\rho_e)-m^2_\mu c^4}}dk\,k^2(k^2+m_\mu^2c^4)^{1/2},
\label{totaled}
\end{align}
and
\begin{align} 
P(\rho,\rho_e,y_p) &= p(\rho,y_p) + \frac{\mu_e^4(\rho_e)}{12\pi^2\hbar^3 c^3}
\nonumber\\
&+ \frac{1}{3\pi^2\hbar^3 c^3}\int_0^{\sqrt{\mu_\mu^2(\rho_e)-m^2_\mu c^4}} 
\frac{dk\,k^4}{(k^2+m_\mu^2 c^4)^{1/2 } },
\label{totalp}
\end{align}
where, $\epsilon(\rho,y_p)$ and $p(\rho,y_p)$ are given in the Eqs.~(\ref{edsk}) and (\ref{prsk}), respectively. The chemical equilibrium and the charge neutrality conditions 
are
\begin{align} 
\mu_n(\rho,y_p) - \mu_p(\rho,y_p) = \mu_e(\rho_e),
\label{mueq}
\end{align}
and
\begin{align}
\rho_p(\rho,y_p) - \rho_e  = \rho_\mu(\rho_e),
\label{charge}
\end{align}
where $\mu_p$ and $\mu_n$ are found from Eq.~(\ref{muqsk}), $\mu_e=\hbar c(3\pi^2\rho_e)^{1/3}$, 
$\rho_p=y_p\rho$, \mbox{$\rho_\mu=[(\mu_\mu^2 - m_\mu^2)^{3/2}]/(3\pi^2\hbar^3 c^3)$}, and 
$\mu_\mu=\mu_e$, for $m_\mu c^2=105.7$~MeV and massless electrons (the ultrarelativistic limit 
is a suitable assumption here since the electron rest mass is around $0.5$~MeV). Regarding the 
muons, their energy density and pressure are given by the last terms of Eqs.~(\ref{totaled}) 
and~(\ref{totalp}), respectively, with the degeneracy factor equal to~$2$. Thus, for each input 
density $\rho$, the quantities $\rho_e$ and $y_p$ are calculated by simultaneously 
solving conditions~(\ref{mueq}) and (\ref{charge}), along with the definitions of 
$\rho_\mu$ and $\mu_\mu$ (both functions of $\rho_e$) previously given in the text.

The properties of a spherically symmetric static neutron star can be studied by taking the energy 
density and pressure as input to the widely known TOV equations, which are given 
by~\cite{tov39,tov39a},
\begin{eqnarray}
\dfrac{dP (r)}{dr} &=& -\dfrac{\left[{\mathcal E} (r) + P (r)\right] \left[m(r) + 4\pi r^3 
P (r)\right]}{r^2\left[1 - \dfrac{2m(r)}{r}\right]} \nonumber \\
\label{tov1}
\end{eqnarray} 
and
\begin{eqnarray}
\dfrac{dm(r)}{dr} &=& 4\pi r^2 {\mathcal E} (r),
\label{tov2}
\end{eqnarray}
where the solution is constrained to the following two conditions at the neutron star 
center: $P(0) = P_c$ (central pressure), and $m(0) = 0$ (central mass). Furthermore, at the star 
surface one has $P(R) = 0$ and $m(R)\equiv M$, with $R$ being the neutron star radius.  
These equations are given in gravitational units, in which $G=1=c$~\cite{glend}. In this specific 
unit system, mass can be expressed in length units and density as inverse length square, as for 
example in km$^{-2}$. On the other hand, in natural units ($\hbar = c = 1$), energy density and 
pressure can also be expressed in fm$^{-4}$, as stated immediately after Eq.~(\ref{prsk}). By mixing 
both, graviational and natural units, one finds $1$~fm$^{-4}=2.6115\times 
10^{-4}$~km$^{-2}$~\cite{glend} that is used to express energy density and pressure in units of 
km$^{-2}$. In this case, mass and radius have the same unit, namely, km (with $1\,M_{\odot} = 
1.4766$~km~\cite{glend}).

\begin{figure}[!htb]
\includegraphics[scale=0.34]{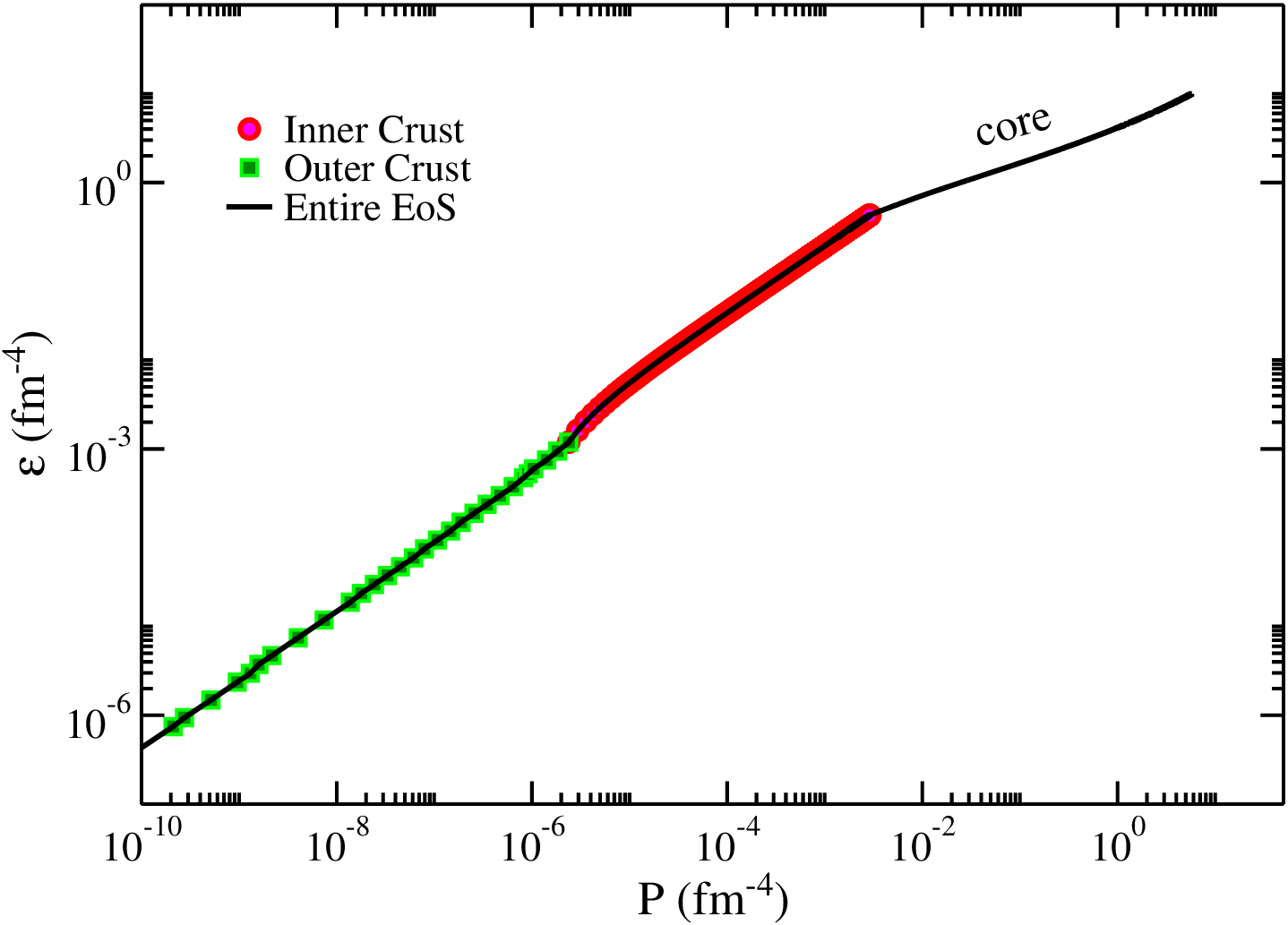}
\caption{Neutron star matter EOS for the KDE0v1 parametrization.}
\label{figkde}
\end{figure}

In order to solve the TOV equations in this work, we take $\mathcal{E}$ and $P$ given in 
Eqs.~(\ref{totaled}) and~(\ref{totalp}) as the EOS that describes the neutron star core. For the 
crust, we consider the different regions defined as the outer and the inner crust. We describe the 
outer crust by the EOS developed by Baym, Pethick and Sutherland (BPS)~\cite{bps} in a density 
region from $\rho=6.3\times10^{-12}\,\mbox{fm}^{-3}$ to $\rho=2.5\times10^{-4}\,\mbox{fm}^{-3}$. 
The exact range is not known, but the outer crust is estimated to exist at densities around 
$10^4$~g/cm$^3$ to $4\times 10^{11}$~g/cm$^3$, i.e, 
around $5\times10^{-12}$~fm$^{-3}$ to $2\times10^{-4}$~fm$^{-3}$, see Refs.~\cite{bps,poly2}, for 
instance. Our choice for this range is compatible with this estimation. A similar range was also 
used in Ref.~\cite{malik19}. For the inner crust region, on the other hand, we impose a polytropic 
form for the total pressure as a function of the total energy density, 
namely, $P(\mathcal{E})=A+B\mathcal{E}^{4/3}$~\cite{poly2,poly1,gogny2}, from 
$\rho=2.5\times10^{-4}\,\mbox{fm}^{-3}$ to $\rho=\rho_t$. Here, $\rho_t$ is related to the 
core-crust transition, in our case estimated from the thermodynamical method described in 
Refs.~\cite{gogny1,cc2,gonzalez19}, for instance. Such a procedure establishes that the 
transition density is defined by the crossing of the EOS with the spinodal section, as one can see 
in Ref.~\cite{debora06}. $A$ and $B$ are constants found by imposing the matching between the outer 
and the inner crust, and between the inner crust and the core. The EOS for the KDEv01 
parametrization is shown in Fig.~\ref{figkde} and illustrates the piecewise structure we use in this 
work.
The remaining CSkP follow the same pattern.

\subsection{Tidal deformability and moment of inertia}

In order to perform a detailed analysis concerning the prediction of the CSkP on the recent GW170817 
event, a very important quantity has to be computed, namely, the tidal deformability. It is one of 
the observed quantities in the binary neutron stars system~\cite{ligo17,ligo18}, which plays a major 
role in constraining hadronic EOS. The induced quadrupole moment $Q_{ij}$ in one neutron star of a 
binary system due to the static external tidal field $E_{ij}$ created by the companion star 
can be written 
as~\cite{tanj10,hind08},
\begin {equation}
Q_{ij} = -\lambda {E}_{ij}. 
\end{equation}
Here, $\lambda$ is the tidal deformability parameter, which can be expressed in terms of 
dimensionless quadrupole tidal Love number $k_2$ as
\begin{equation}
\label{tidal}
\lambda= \frac{2}{3} {k_2}R^{5}.
\end{equation}
The dimensionless tidal deformability $\Lambda$ (i.e., the dimensionless version of 
$\lambda$) is connected with the compactness parameter $C= M/R$ through 
\begin{equation}
\Lambda= \frac{2k_2}{3C^5}. 
\label{dtidal}
\end{equation}  

The tidal Love number $k_2$ is obtained as
\begin{align}
k_2 &=\frac{8C^5}{5}(1-2C)^2[2+2C(y_R-1)-y_R]\nonumber\\
&\times\Big\{2C [6-3y_R+3C(5y_R-8)] \nonumber\\
&+ 4C^3[13-11y_R+C(3y_R-2) + 2C^2(1+y_R)]\nonumber\\
&+ 3(1-2C)^2[2-y_R+2C(y_R-1)]{\rm ln}(1-2C)\Big\}^{-1},\,\,
\label{k2}
\end{align}
with $y_R\equiv y(R)$, where $y(r)$ is found from the solution of
\begin{equation}
r \frac{dy}{dr} + y^2 + yF(r) + r^2Q(r)=0,
\label{ydef}
\end{equation}
with
\begin{equation}
F(r) = \frac{r - 4\pi r^3[\mathcal{E}(r) - P(r)]}{r - 2m(r)}
\end{equation}
and
\begin{align}
Q(r)&=\frac{4\pi r\left[5\mathcal{E}(r) + 9P(r) + \frac{\mathcal{E}(r)+P(r)}{\partial 
P(r)/\partial\mathcal{E}(r)}-\frac{6}{4\pi r^2}\right]}{r - 2m(r)} 
\nonumber\\ 
&- 4\left[ \frac{m(r)+4\pi r^3 P(r)}{r^2(1-2m(r)/r)} \right]^2.
\end{align}
In order to find $y(r)$,  Eq.~(\ref{ydef}) has to be solved as part of a coupled system 
containing the TOV equations given in Eqs.~(\ref{tov1}) and (\ref{tov2}).

The dimensionless tidal deformabilities of a binary neutron stars system, namely, 
$\Lambda_1$ and $\Lambda_2$, can be combined to yield the weighted average 
as~\cite{ligo17} 
\begin{align}
\tilde{\Lambda} = \frac{16}{13}\frac{(m_1+12m_2)m_1^4\Lambda_1 + 
(m_2+12m_1)m_2^4\Lambda_2}{(m_1+m_2)^5},
\label{Lambdatilde}
\end{align}
where $m_1$ and $m_2$ are masses of the two companion stars.

Finally, in order to verify whether the $I$-Love relation also applies to the CSkP, we solve the 
Hartle's slow rotation equation given in Refs.~\cite{phil18,hartle,yagi13}, namely,
\begin{eqnarray}
0&=&[r-2m(r)]\frac{d^2\omega}{dr^2} - 16\pi r[\mathcal{E}(r)+P(r)]\omega(r)
\nonumber\\
&+& 4\left\{\left(1-\frac{2m(r)}{r}\right) - \pi 
r^2[\mathcal{E}(r)+P(r)]\right\}\frac{d\omega}{dr},
\label{hartle}
\end{eqnarray}
coupled to the TOV equations. Since the binary system related to the GW170817 event rotates 
slowly, according to Ref.~\cite{phil18}, the Hartle's method can be safely used.

From the solution $\omega(r)$, one determines the moment of inertia 
through the relation $I=R^3(1-\omega_R)/2$, with $\omega_R\equiv \omega(R)$. The dimensionless 
version of this quantity is defined as $\bar{I}\equiv I/M^3$. 
The linearized version of Eq.~(\ref{hartle}) is given by
\begin{equation}
r \frac{d\zeta}{dr} + \zeta \left(\zeta + 3 \right) - \frac{4\pi r^2[\mathcal{E}(r)+P(r)](\zeta + 
4)}{1 - 2 m(r)/r}=0
\label{lin1}
\end{equation}
with
\begin{equation}
\zeta \equiv \frac{1}{\omega} \frac{d\omega}{dr}.
\label{lin2}
\end{equation}
In this case, the boundary conditions are $\zeta = 0$ at the center, and $I =[\zeta/(3+\zeta)] 
R^3/2$ at the surface. This formulation is easier to be numerically integrated since it is a 
first-order differential equation.

\section{Results and Discussions}
\label{result}

\begin{figure}[!htb]
\includegraphics[scale=0.34]{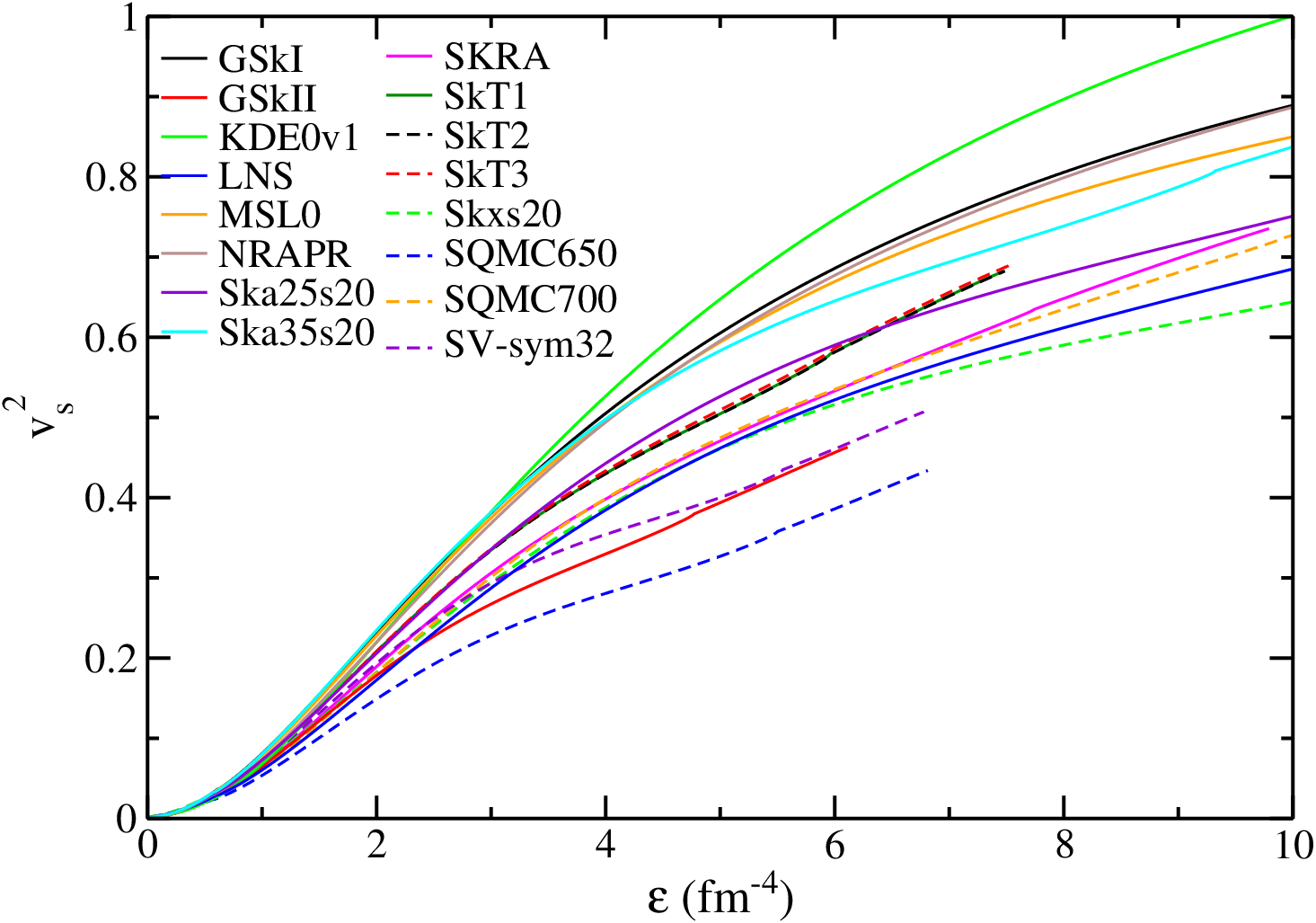}
\caption{Squared sound velocity as a function of total energy density for the CSkP.}
\label{sound}
\end{figure}

As all the CSkP come from a nonrelativistic mean field model, at zero 
temperature regime, the causal limit may be broken at the high density region, since the 
sound velocity ($v_s$) increases with density, or equivalently, with energy density. 
However, for the CSkP we verify that $v_s^2=\partial P/\partial\mathcal{E}$ 
exceeds $c^2=1$ only at very high energy density 
values, as we can see in Fig.~\ref{sound}.

From this figure, one can verify that the CSkP obey the causal limit up to a range of 
$\mathcal{E}\lesssim 10$~fm$^{-4}$. By comparing these results with those obtained for relativistic 
mean-field (RMF) parametrizations in Fig. 2 of Ref.~\cite{dutra16}, a clear difference in behavior 
is observed. The RMF parametrizations present a saturation for the sound velocity unlike the Skyrme 
ones, that always increase. Despite this increasing dependence, Fig.~\ref{sound} shows that it is 
possible to describe neutron star matter with CSkP within a particular range of energy densities. 
The description of global properties of neutron stars by other different Skyrme 
parametrizations can be found, for instance, in Refs.~\cite{phil18,malik18,sk1,sk2,sk3,sk4,sk5,sk6}. 
Notice that the curves that end below an energy density around 8 fm$^{-4}$ refer to models that 
stop converging at these lower densities. Had we plotted the sound velocity as a function of the 
baryonic density, the behavior would be similar. For all models, the baryonic density corresponding 
to the energy density equal to 2 (6) fm$^{-4}$ is of the order of 0.4 (1.0) fm$^{-3}$ and a ratio of 
2.4 (6) times their saturation densities.
Furthermore, other nonrelativistic models such as Gogny, Simple Effective Interaction (SEI), and 
momentum-dependent interaction (MDI), based on finite range interactions unlike the Skyrme model, 
are also used in neutron star calculations, see Refs.~\cite{gogny2,gogny1,sei1,sei2,kras19}.

\begin{figure}[!htb]
\includegraphics[scale=0.34]{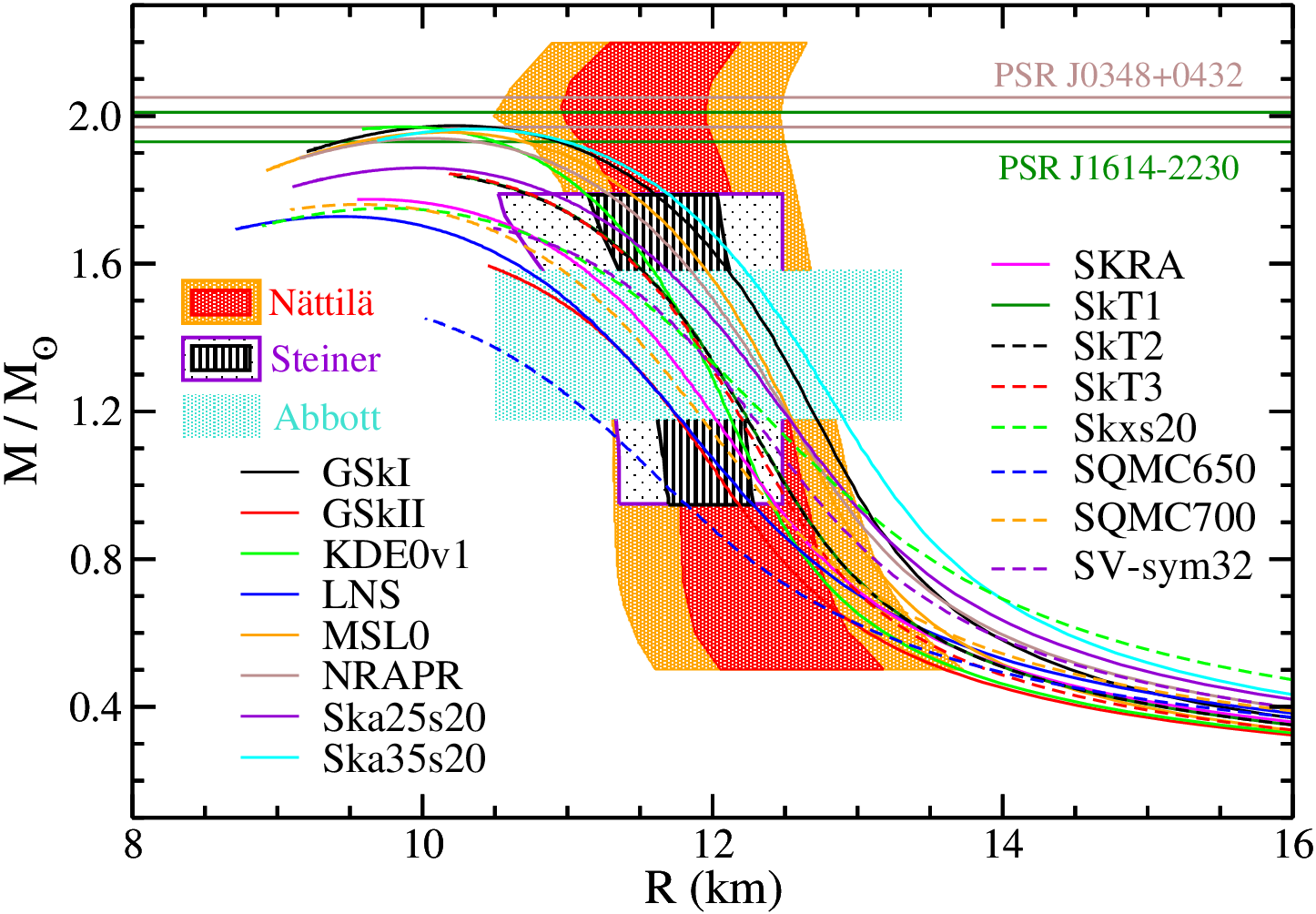}
\caption{Neutron star mass-radius profiles for the CSkP. Horizontal bands indicate the 
masses of PSR~J1614-2230~\cite{nature467-2010} and PSR~J0348+0432~\cite{science340-2013}. Turquoise 
band: limits from the GW170817 event found in Ref.~\cite{ligo18}.}
\label{massradius}
\end{figure}

The mass-radius profiles predicted by the CSkP are shown in Fig.~\ref{massradius}. In this figure, 
horizontal bands in magenta and green colors indicate respectively the observational data of pulsar 
masses of PSR~J1614-2230~\cite{nature467-2010} and PSR~J0348+0432~\cite{science340-2013}. We also 
show the empirical constraints for the mass-radius profile for the cold dense matter inside the 
neutron star. They were obtained from a Bayesian analysis of type-I x-ray burst observations by 
N\"attil\"a, {\it et al.} in Ref.~\cite{nat16} (outer orange and inner red bands), and from a 
mass-radius coming from six sources, namely, three from transient low-mass x-ray binaries and three 
from type-I x-ray bursts with photospheric radius expansion, by Steiner {\it et al.} in 
Ref.~\cite{stein10} (outer white and inner black bands). In the same figure, it is also represented 
by the turquoise band the region of masses and radii obtained from the analysis of the GW170817 
event regarding the binary neutron star system~\cite{ligo18}. 

These observations imply that the neutron star mass predicted by any theoretical model should reach 
the limit of $M\sim 2.0M_{\odot}$. From the results, we find that the maximum masses obtained by 
the GSkI, Ska35s20, MSL0, NRAPR, and KDE0v1 parametrizations are in agreement with at least 
one of these boundaries~\cite{nature467-2010,science340-2013}. Very recently, another massive 
millisecond pulsar was confirmed, namely, MSP J0740+6620, with a mass of $2.14^{+0.20}_{-0.18} 
M_\odot$ within 95.4\% credibility or $2.14^{+0.10}_{-0.09} M_\odot$ within 68.3\% credibility 
\cite{nature_2019}. Indeed, four of the models mentioned above, namely, GSkI, Ska35s20, 
MSL0 and KDE0v1, also lie within the former mass limit of this pulsar ($2.14^{+0.20}_{-0.18} 
M_\odot$), as one can verify from the results presented in Table~\ref{tab1}.
\begin{table*}[!htb]
\centering
\caption{Transitions values ($\rho_t$, $\epsilon_t$, and $p_t$) along with the stellar matter 
properties obtained from the CSkP: maximum neutron stars mass ($M_{\rm max}$) and its corresponding 
radius ($R_{\rm max}$), compactness ($C_{\rm max}$), and central energy density ($\mathcal{E}_c$) 
along with the radius ($R_{1.4}$) and compactness ($C_{1.4}$) of the canonical star.}
\begin{tabular}{|lccccccccc|}
\hline
Parameter & $\rho_t$ & $\epsilon_t$ & $p_t$ &$M_{\rm max}$ & $R_{\rm max} $ & $C_{\rm max}$ & 
$\mathcal{E}_c$ & $R_{\rm 1.4}$ & $C_{\rm 1.4}$ 
\\
& (fm$^{-3}$) & (fm$^{-4}$) & (MeV/fm$^3$)  & ($M_\odot$) & (${\rm km}$) &
($M_\odot/{\rm km}$) & (${\rm fm}^{-4}$) & (${\rm km}$) & ($M_\odot/{\rm km}$)\\ 
\hline
GSkI & 0.081 & 0.390 & 0.492 & 1.974 & 10.229 & 0.193 & 8.095 & 12.419 & 0.113\\ 
GSkII & 0.087 & 0.417 & 0.532 & 1.594 & 10.452 & 0.153 & 6.094 & 11.267 & 0.124\\ 
KDE0v1 & 0.089 & 0.429 & 0.570 & 1.970 & 9.863 & 0.200 & 8.573 & 11.858 & 0.118\\ 
LNS & 0.087 & 0.417 & 0.612 & 1.728 & 9.436 & 0.183 & 9.848 & 11.278 & 0.124 \\ 
MSL0 & 0.079 & 0.381 & 0.441 & 1.956 & 10.155 & 0.193 & 8.198 & 12.258 & 0.114\\ 
NRAPR & 0.083 & 0.397 & 0.553 & 1.939 & 10.032 & 0.193 & 8.481 & 12.188 & 0.115\\ 
Ska25s20 & 0.083 & 0.399 & 0.577 & 1.859 & 9.981 & 0.186 & 8.711 & 12.112 & 0.116 \\ 
Ska35s20 & 0.085 & 0.406 & 0.594 & 1.964 & 10.358 & 0.190 & 7.975 & 12.553 & 0.112 \\
SkRA & 0.083 & 0.398 & 0.529 & 1.774 & 9.643 & 0.184 & 9.332 & 11.599 & 0.121 \\ 
SkT1 & 0.087 & 0.420 & 0.560 & 1.838 & 10.215 & 0.180 & 7.458 & 11.895 & 0.118\\ 
SkT2 & 0.087 & 0.419 & 0.560 & 1.837 & 10.210 & 0.180 & 7.467 & 11.892 & 0.118\\ 
SkT3 & 0.087  & 0.418 & 0.541 & 1.844 & 10.185 & 0.181 & 7.536 & 11.870 & 0.118\\ 
Skxs20 & 0.081  & 0.388 & 0.615 & 1.750 & 9.771 & 0.179 & 9.306 & 11.815 & 0.118\\
SQMC650 & 0.093 & 0.446 & 0.694 & 1.452 & 10.029 & 0.145 & 6.790 & 10.355 & 0.135 \\
SQMC700 & 0.088 & 0.422 & 0.630 & 1.760 & 9.568 & 0.184 & 9.520 & 11.473 & 0.122 \\ 
SV-sym32 & 0.085 & 0.410 & 0.589 & 1.696 & 10.490 & 0.162 & 6.760 & 11.819 & 0.118 \\ 
\hline
\end{tabular}
\label{tab1}
\end{table*}

Furthermore, the radii obtained from these 
parametrizations for the canonical star of $M=1.4M_\odot$ are also inside the bands calculated in 
Refs.~\cite{nat16,stein10}. The remaining CSkP underestimate the observed data regarding the 
neutron star mass. Finally, concerning the GW170817 constraint, one can verify that all the CSkP 
are entirely compatible with this particular restriction. The exception is the SQMC650 
parametrization, that satisfies the constraint only partially.

In Table~\ref{tab1} we also present some properties regarding the CSkP, namely, the transition 
point (transition density, energy density and pressure) found by the thermodynamical 
method~\cite{gogny1,cc2,gonzalez19}, and neutron star matter quantities. For the latter, we show the 
maximum neutron star mass and corresponding radius, compactness and central energy density. We also 
tabulate the radius and compactness related to the canonical neutron star. It is worth mentioning 
that the central energy density of all CSkP are compatible with the causal limit, as one can verify 
from Fig.~\ref{sound}.

In the recent literature, a lot of effort has been put to 
constrain the radius of the canonical 
neutron star, see for instance, Refs.~\cite{malik18,yeun18,elia18,zhan19,caro18,tews18}. In 
Ref.~\cite{malik18}, Tuhin Malik {\it et al.} have discussed this constraint by using Skyrme and 
RMF models and their calculations suggest the range of $11.82~\mbox{km} \leqslant R_{1.4}\leqslant 
13.72~\mbox{km}$. By using a set of more realistic models and the neutron skin values as a new 
constraint, F. J. Fattoyev {\it et al.} have shown the upper limit for $R_{1.4}$ as 
$13.76~\mbox{km}$~\cite{fatt18}. In Ref.~\cite{yeun18}, Yeunhwan Lim {\it et al.} have used chiral 
effective field theory and constraints from nuclear experiments to establish the range of 
$10.36~\mbox{km}\leqslant R_{1.4} \leqslant 12.87~\mbox{km}$. Elias R. Most {\it et al.} have 
studied the constraint on $R_{1.4}$ with a large number of EOS with pure hadronic matter without any 
kind of phase transition~\cite{elia18}. They found the value of $R_{1.4}$ inside the range of 
$12.00~\mbox{km} \leqslant R_{1.4} \leqslant 13.45~\mbox{km}$, with the most likely value of 
$R_{1.4}=12.39~\mbox{km}$. From the above discussion, we can estimate an specific 
range for $R_{1.4}$ encompassing the previous ones as $10.36~\mbox{km} \leqslant R_{1.4} \leqslant 
13.76~\mbox{km}$. Our calculations for $R_{1.4}$ from the CSkP show a minimum value of 10.36~km 
(SQMC650 parametrization), while the maximum value is given by 12.55~km (Ska35s20 parameter set). 
Both maximum and minimum values present very good agreement with the composite range. As a 
consequence, the 5 CSkP predicting neutron star mass around two solar masses, namely, GSkI, KDE0v1, 
MSL0, NRAPR, and Ska35s20, also present $R_{1.4}$ compatible with the range mentioned above. The 
minimum value of this quantity is obtained by the KDE0v1 parametrization: 
$R_{1.4}=11.86$~km, while the maximum value is found by the Ska35s20 set, namely, 
$R_{1.4}=12.55$~km. This number is close to the 
most likely value of $R_{1.4}$ given in Ref.~\cite{elia18}, namely, $R_{1.4}=12.39~\mbox{km}$.
\begin{figure}[!htb]
\includegraphics[scale=0.34]{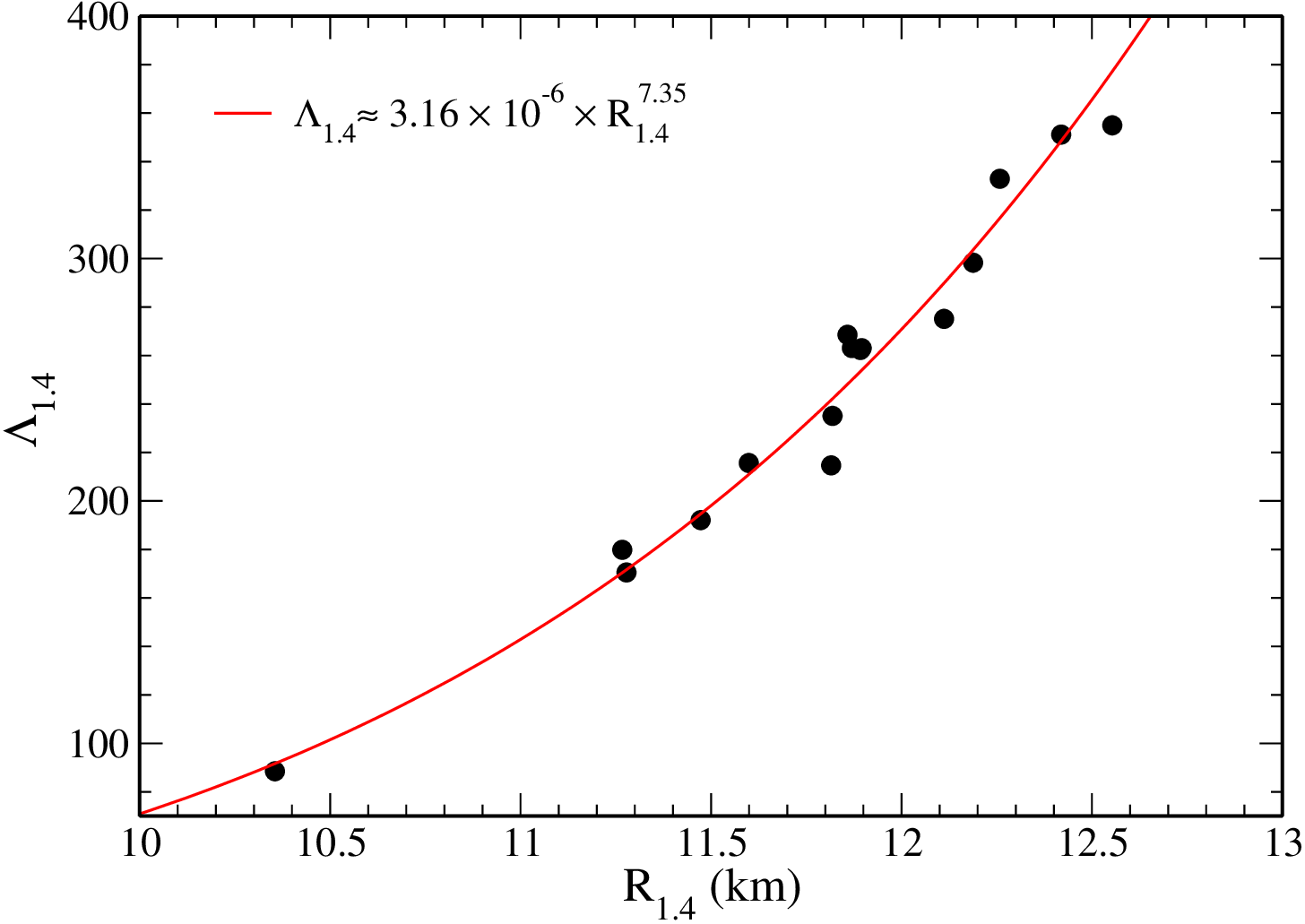}
\caption{Canonical neutron star tidal deformability as a function of its radius 
for the CSkP. Solid line: fitting curve.}
\label{canonical}
\end{figure}

In searching for other possible correlations in the context of the neutron star binary system, one 
can notice from Eq.~(\ref{dtidal}) that $\Lambda\propto R^5$ is not a good assumption, since the 
tidal Love number $k_2$ depends on the neutron star radius in a nontrivial way, as seen in 
Eq.~(\ref{k2}). In this context, we try to find a correlation between the radius and tidal 
deformability for the CSkP for the canonical star, the one with $M=1.4M_{\odot}$. The obtained 
results for $\Lambda_{1.4}$ as a function of $R_{1.4}$ are shown in Fig.~\ref{canonical}, with a 
similar qualitative behavior in comparison with the study performed in Ref.~\cite{tsang}, for 
instance. From the points shown in the figure, we could establish a fitting curve correlating 
$\Lambda_{1.4}$ as a function of $R_{1.4}$, namely, $\Lambda_{1.4} \approx 3.16 \times 
10^{-6}R_{1.4}^{7.35}$. This correlation presents different numbers in comparison with those found 
from predictions of EOS constructed by chiral effective field theory at low densities and the 
perturbative QCD at very high baryon densities using polytropes~\cite{anna18}, several energy 
density functional within RMF models~\cite{fatt18}, and both RMF and Skyrme Hartree-Fock energy 
density functionals~\cite{malik18}. In these cited works, the authors found $\Lambda_{1.4} \approx 
2.88 \times 10^{-6}R_{1.4}^{7.5}$~\cite{anna18}, $\Lambda_{1.4}\approx 7.76 \times 
10^{-4}R_{1.4}^{5.28}$~\cite{fatt18}, and $\Lambda_{1.4}\approx 9.11 \times 
10^{-5}R_{1.4}^{6.13}$~\cite{malik18}. A recent analysis performed in Ref.~\cite{rmfdef} by using a 
set of consistent RMF parametrizations, pointed out to $\Lambda_{1.4}\approx 2.65 \times 
10^{-5}R_{1.4}^{6.58}$. These different fittings point towards the non-existence of an universal 
power-law  of $\lambda$ or $\Lambda$ as a function of $R$, as one could naively think by looking at 
Eqs.~(\ref{tidal}) and~(\ref{dtidal}).

\begin{figure}[!htb]
\includegraphics[scale=0.34]{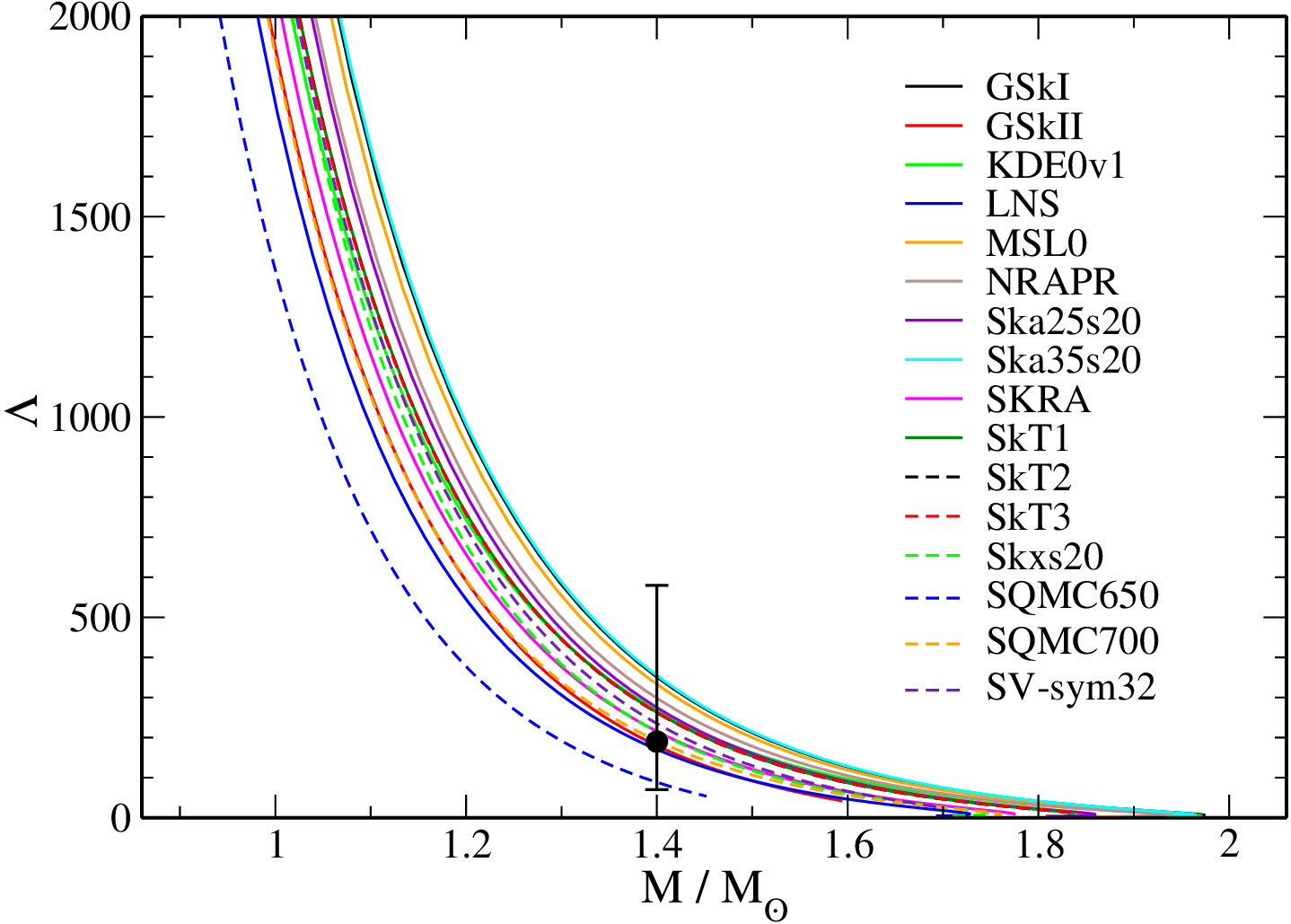}
\caption{$\Lambda$ as a function of $M$ for the CSkP. Full circle: recent result of 
$\Lambda_{1.4}=190_{-120}^{+390}$ obtained by LIGO an Virgo Collaboration~\cite{ligo18} 
related to the canonical star.}
\label{lambdamass}
\end{figure}

For the sake of completeness, in Fig.~\ref{lambdamass} we plot the dimensionless tidal 
deformability $\Lambda$ of a static neutron star as a function of its mass for the CSkP. 
The tidal deformability decreases nonlinearly with the neutron star mass for all 
parametrizations. At $M=1.4M_{\odot}$, the resulting values of $\Lambda$ stand within a 
range of around $100-350$ for the CSkP, which are within the upper limit of 
$\Lambda_{1.4}\leqslant 800$ of LIGO + Virgo gravitational detection~\cite{ligo17}, and 
also the recent updated range of $\Lambda_{1.4}=190_{-120}^{+390}$~\cite{ligo18}.

\begin{figure}[!htb]
\includegraphics[scale=0.34]{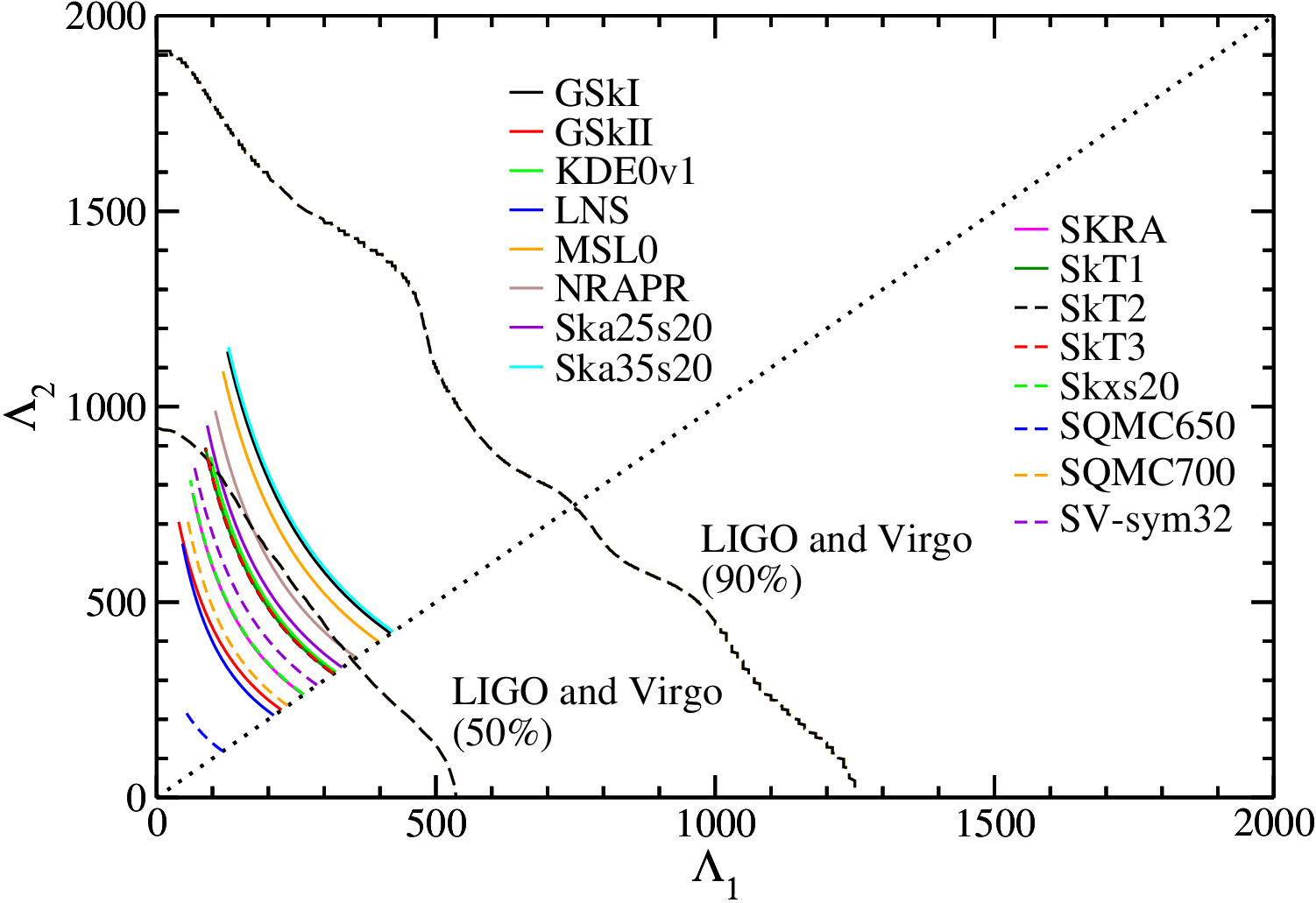}
\caption{Tidal deformability parameters predicted by the CSkP for the case of high-mass
($\Lambda_1$) and low-mass ($\Lambda_2$) components of the observed GW170817 event. The 
90\% and 50\% confidence lines were taken from recent findings of Ref.~\cite{ligo18}.}
\label{l1l2}
\end{figure}
In Fig.~\ref{l1l2} we plot the tidal deformabilities $\Lambda_1$ and $\Lambda_2$ of the binary 
neutron stars system with component masses of $m_1$ and $m_2$ ($m_1>m_2$). The diagonal dotted line 
corresponds to the $\Lambda_1=\Lambda_2$ case in which $m_1=m_2$. The analysis takes into account 
the range for $m_1$ given by $1.365\leqslant m_1/M_\odot \leqslant 1.60$, as pointed out in 
Ref.~\cite{ligo17}. The mass of the companion star, $m_2$, is calculated through the relationship 
between $m_1$, $m_2$ and the chirp mass given by
\begin{align}
\mathcal{M}_c = \frac{(m_1m_2)^{3/5}}{(m_1+m_2)^{1/5}}.
\end{align}
In this equation, $\mathcal{M}_c$ is fixed at the observed value of $1.188M_\odot$~\cite{ligo17}. 
The upper and lower dash lines correspond to the 90\% and 50\% confidence limits respectively, which 
are obtained from the recent analysis of the GW170817 event~\cite{ligo18}. This figure shows that 
all $16$ CSkP are completely inside the 90\% credible region predicted by the GW170817 
data~\cite{ligo18}. Other kind of models, such as some relativistic ones~\cite{rmfdef,apj} also 
present good agreement with this particular region predicted by the the LIGO and Virgo 
Collaboration.

We also calculate the ranges related to the mass weighted tidal deformability as defined in 
Eq.~(\ref{Lambdatilde}) by using $\mathcal{M}_c=1.188M_\odot$ and the aforementioned variations of 
$m_1$ and $m_2$. The results are presented in Table~\ref{tabltilde}.
\begin{table}[!htb]
\centering
\caption{Ranges for $\tilde{\Lambda}$ predicted by the CSkP.}
\begin{tabular}{lc}
\hline
Parameter & Range of $\tilde{\Lambda}$  \\
\hline
GSkI & 420 - 427\\ 
GSkII & 224 - 238\\ 
KDE0v1 & 321 - 326\\ 
LNS & 210 - 223\\ 
MSL0 & 398 - 406\\ 
NRAPR & 358 - 366\\ 
Ska25s20 & 332 - 344\\ 
Ska35s20 & 424 - 431\\
SkRA &  263 - 274\\ 
SkT1 & 317 - 325\\ 
SkT2 & 316 - 325\\ 
SkT3 & 318 - 325\\ 
Skxs20 & 264 - 281\\
SQMC650 & 119 - 120\\
SQMC700 &  235 - 247\\ 
SV-sym32 &  288 - 298\\ 
\hline
\end{tabular}
\label{tabltilde}
\end{table}

One can see that all the CSkP present $\tilde{\Lambda}$ in full agreement with the range determined 
in Ref.~\cite{ligo17}, namely, $\tilde{\Lambda}\leqslant 800$, when the chirp mass given by 
$\mathcal{M}_c=1.188M_\odot$~\cite{ligo17} is used. Furthermore, if we use the value of 
$\mathcal{M}_c=1.186M_\odot$~\cite{ligo19}, the results presented in Table~\ref{tabltilde} change 
only slightly. For instance, for the GSkI parametrization the range changes to $424\leqslant 
\tilde{\Lambda}\leqslant 432$.

Regarding the calculation of deformabilities, it is worth mentioning that the inner crust-core phase 
transition may be slightly different if obtained from the thermodynamical, dynamical approximations 
or from the interface between the pasta and the homogeneous phases, as can be seen in 
Ref.~\cite{PRC035804-2009}. Moreover, in Ref.~\cite{poly2} it is claimed that the inner
crust does not play an important role in the calculation of the deformability, what was 
corroborated in Ref.~\cite{nosso_QMC}, where the pasta phase was explicitly taken into account to 
describe the inner crust. Hence, the differences found by different models due to the use of 
another prescription (dynamical instead of thermodynamical method) would be consistent and  would 
certainly lead to the same conclusions.

\begin{figure}[!htb]
\includegraphics[scale=0.34]{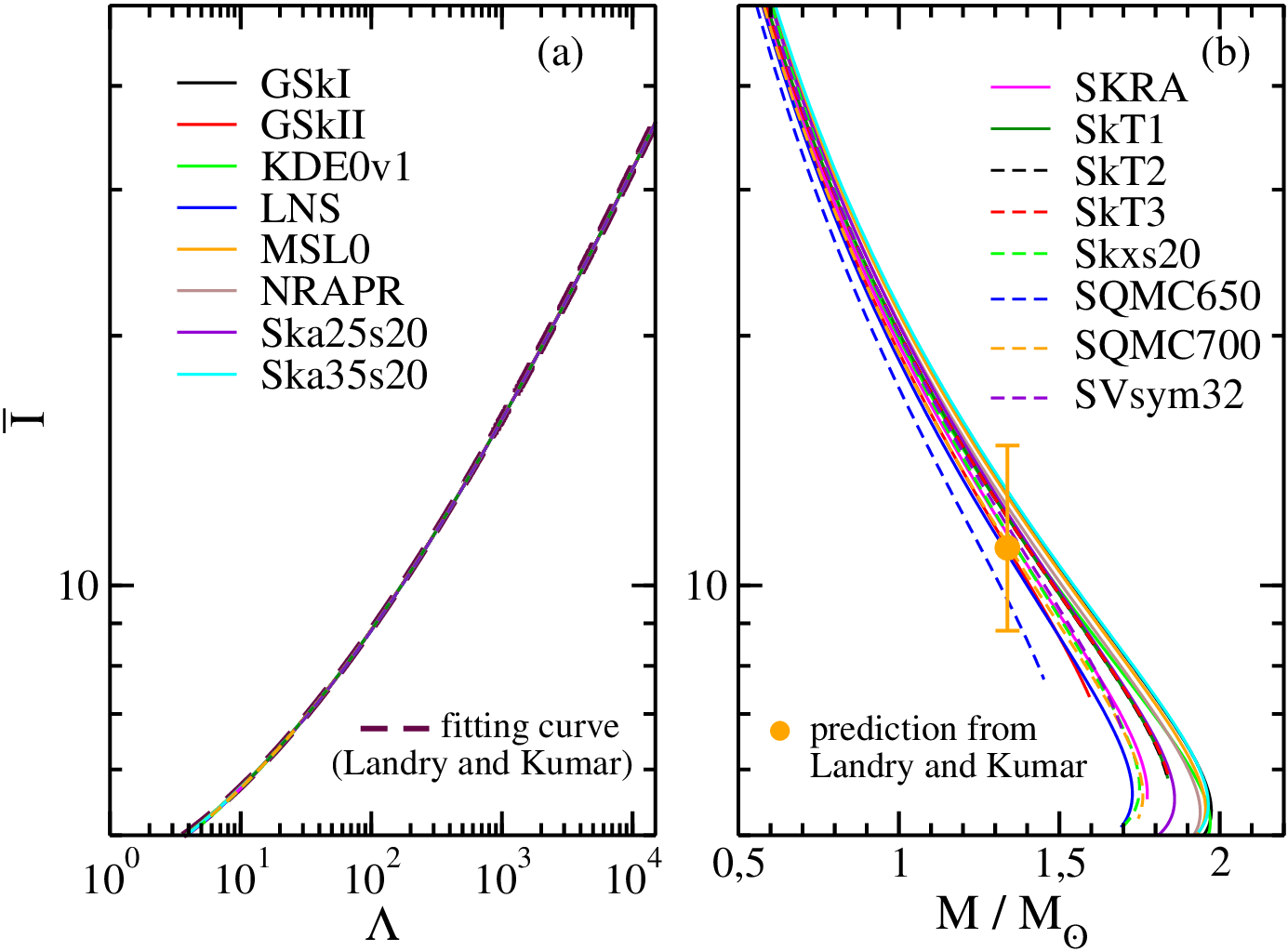}
\caption{$\bar{I}$ as a function of (a)~$\Lambda$, and (b)~$M/M_\odot$. The same parametrizations 
are used in both panels. Dashed orange curve: fitting curve of Ref.~\cite{phil18}. Orange circle: 
predictions from Ref.~\cite{phil18} for the dimensionless moment of inertia of the \mbox{PSR 
J0737-3039} primary component pulsar.}
\label{inertia}
\end{figure}

Finally, we show in Fig.~\ref{inertia} the dimensionless moment of inertia calculated from the 
CSkP. Since in our calculations Eq.~(\ref{hartle}), or Eqs.~(\ref{lin1}) and (\ref{lin2}) are 
solved coupled to the TOV equations and also to Eq.~(\ref{ydef}), one can simultaneously extract 
information regarding $M$, $\Lambda$ and $I$ (or $\bar{I}$). In panel (a), in which we show 
$\bar{I}$ as a function of $\Lambda$, it is verified that all CSkP are indistinguishable. This 
universality is known as the \mbox{$I$-Love} relation. In Ref.~\cite{phil18}, this feature was 
obtained for parametrizations coming from the relativistic mean-field model and the Skyrme one. For 
the latter, among the 24 Skyrme parametrizations employed in Ref.~\cite{phil18}, only 
one is also included in the set we have employed in the present work, namely, the KDE0v1 
parametrization. Here, we confirm the universal behavior of the $\bar{I}\times\Lambda$ curves for 
all the CSkP. The fitting curve generated in Ref.~\cite{phil18} is also shown in 
Fig.~\ref{inertia}{\color{blue}a}.

In Ref.~\cite{phil18}, it was also obtained the range of 
$\bar{I}_\star\equiv\bar{I}(M_\star)=11.10^{+3.64}_{-2.28}$ for the \mbox{PSR J0737-3039} primary 
component pulsar with mass $M_\star=1.338M_\odot$. Such  numbers were determined through the 
relation between $\Lambda_\star\equiv\Lambda(M_\star)$ and $\Lambda_{1.4}$, known as the 
\mbox{binary-Love} relation. The combination between the \mbox{$I$-Love} and the \mbox{binary-Love} 
relations, along with the GW170817 constraint of $\Lambda_{1.4}=190^{+390}_{-120}$ coming from the 
LIGO and Virgo Collaboration, allowed the authors to establish the limits given by 
$\bar{I}_\star=11.10^{+3.64}_{-2.28}$. As one can see in Fig.~\ref{inertia}{\color{blue}b}, the CSkP 
present $9.62 \leqslant \bar{I}_\star \leqslant 13.01$, values which lie inside the predicted 
range.

\section{Summary and Conclusions}
\label{summary}

In this paper we have revisited the Skyrme parametrizations that were shown to satisfy 
several nuclear matter constraints in Ref.~\cite{dutra12}, named as the consistent Skyrme 
parametrizations (CSkP), and confronted them with astrophysical constraints and predictions on the 
GW170817 event studied by LIGO and Virgo Collaboration in recent papers~\cite{ligo17,ligo18}. 
Concerning the applicability of these nonrelativistic models at the high density regime of the 
stellar matter, we have shown that causality is not broken at the energy density range of interest, 
as one can see from Fig.~\ref{sound}, and from the comparison with the central energy density 
obtained from  the CSkP and presented in Table~\ref{tab1}. Our calculations also pointed out to a 
radius range of $\mbox{10.36~km} \leqslant R_{1.4} \leqslant \mbox{12.55~km}$ according to the 
predictions of the CSkP. It was also shown that only the GSkI, KDE0v1, MSL0, NRAPR, and Ska35s20 
parametrizations are able to produce neutron stars with mass around $2M_\odot$, value established 
form observational analysis of PSR~J1614-2230~\cite{nature467-2010} and 
PSR~J0348+0432~\cite{science340-2013} pulsars. They also establish the more stringent range of 
$11.86~\mbox{km} \leqslant R_{1.4} \leqslant 12.55~\mbox{km}$ for the canonical star radius. This 
range is similar to the one found in Ref.~\cite{tsang19}, namely, $11~\mbox{km} \leqslant R_{1.4} 
\leqslant 12~\mbox{km}$, in which the authors analyzed 5 out of more than two hundred 
Skyrme parametrizations also investigated in the work. In their study, they also used a piecewise 
way to construct the EOS, namely, the BPS equation for the outer crust, the same form of 
the polytropic equation of state as in the present work ($P=A+B\mathcal{E}^{4/3}$), the Skyrme 
model for the core, and finally, another polytropic form ($P\sim \rho^\gamma$) for the density 
region above $3\rho_0$.

Concerning the predictions of the CSkP on the recent GW170817 event, it was shown that all CSkP, 
except the SQMC650 one, present a mass-radius profile in full agreement with the constraint region 
given in Ref.~\cite{ligo18}. Furthermore, by investigating the results regarding the canonical star 
($M=1.4M_\odot$), our results pointed out to a correlation given by $\Lambda_{1.4} \approx 3.16 
\times 10^{-6}R_{1.4}^{7.35}$ between the dimensionless tidal deformability and the radius. From 
this correlation, we found that the CSkP present values of $\Lambda_{1.4}$ completely inside the 
ranges of $\Lambda_{1.4}\leqslant 800$~\cite{ligo17} and even the recent one given by 
$\Lambda_{1.4}=190_{-120}^{+390}$~\cite{ligo18}, as one can see in Figs.~\ref{canonical} 
and~\ref{lambdamass}. We  have also calculated the dimensionless tidal deformabilities of the 
binary neutron stars system, $\Lambda_1$ and $\Lambda_2$ (see Fig.~\ref{l1l2}), and found that the 
CSkP are completely inside the region defined by 90\% credible region in the 
$\Lambda_1\times\Lambda_2$ graph, predicted by the recent paper from LIGO and Virgo 
Collaboration~\cite{ligo18}. In addition, we verified that the prediction presented in 
Ref.~\cite{phil18} on the dimensionless moment of inertia for the \mbox{PSR J0737-3039} pulsar, 
namely, $\bar{I}_\star=11.10^{+3.64}_{-2.28}$, is attained by the CSkP.

As a last comment, we mention that our study refers only to the predictions of the CSkP with their 
equations of state given in Eqs.~(\ref{edsk})-(\ref{muqsk}), i. e., no hyperons and no hadron-quark 
phase transitions are considered. Specifically for the first treatment, the interactions between 
hyperons and nucleons and between hyperons themselves, which are unknown at present, can be modeled 
in different ways, see for instance, Refs.~\cite{dutra16,sk5}. For the latter, there is not a 
unique model that effectively mimics QCD, which makes this study also model dependent from the quark 
matter considerations. A more detailed and complete study of how these treatments can affect the 
deformability calculations will be addressed in future works.

\section*{ACKNOWLEDGMENTS}
This work is a part of the project INCT-FNA Proc. No. 464898/2014-5, partially supported by 
Conselho Nacional de Desenvolvimento Cient\'ifico e Tecnol\'ogico (CNPq) under grants 301155/2017-8 
(D. P. M.), 310242/2017-7 and 406958/2018-1 (O. L.) and 433369/2018-3 (M. D.), by Funda\c{c}\~ao de 
Amparo \`a Pesquisa do Estado de S\~ao Paulo (FAPESP) under thematic projects No. 2013/26258-4 
(O. L.), 2017/05660-0 (O. L., M. D., M. B.), 2014/26195-5 (M. B.), and National key R\&D 
Program of China, grant No. 2018YFA0404402 (S. K. B.).


\end{document}